\documentclass[a4paper,UKenglish,cleveref, autoref, thm-restate]{lipics-v2021}

\nolinenumbers

\usepackage{subfiles}
\usepackage{standalone}
\usepackage[utf8]{inputenc}
\usepackage{amsthm}
\usepackage{amsmath}
\usepackage{algorithm}
\usepackage{mathtools}
\usepackage{url}
\usepackage{empheq}
\usepackage[noend]{algpseudocode}
\usepackage{enumitem}
\usepackage{graphicx}
\usepackage{caption}
\usepackage{subcaption}
\usepackage{todonotes}

\algrenewcommand\algorithmicloop{\textbf{repeat}}
\algrenewcommand\algorithmicdo{\textbf{}}
\algrenewcommand\algorithmicend{\textbf{}}
\algnewcommand\algorithmicforeach{\textbf{for each}}
\algdef{S}[FOR]{ForEach}[1]{\algorithmicforeach\ #1\ \algorithmicdo}

\usepackage{xcolor}
\usepackage{mathtools} 

\DeclarePairedDelimiter{\floor}{\lfloor}{\rfloor}

\newtheoremstyle{case}{}{}{}{}{}{:}{ }{}
\theoremstyle{case}

\newcommand{\defn}[1]{\emph{\textbf{#1}}}
\newcommand{\id}[1]        {\ifmmode\mathit{#1}\else\textit{#1}\fi}

\newcommand{\ceil}[1]{\left\lceil#1\right\rceil}

\newcommand{\secref}[1]{Section~\ref{sec:#1}}

\newcommand{\etal}{\textit{et al.}\@}

\newcommand{\Des}{\id{Des}}
\newcommand{\Anc}{\id{Anc}}

\newcommand{\sfB}{\mathsf{B}}
\newcommand{\sfC}{\mathsf{C}}

\DeclareMathOperator*{\union}{\bigcup}
\newcommand{\triples}{\mathsf{triples}}

\newcommand{\parent}{\mathsf{par}}

\title{Nested Active-Time Scheduling}  

\author{Nairen Cao}{Georgetown University, Washington D.C., USA}{nairen@ir.cs.georgetown.edu}{}{}

\author{Jeremy T. Fineman}
{Georgetown University, Washington D.C., USA}
{jfineman@cs.georgetown.edu}
{}
{}

\author{Shi Li}
{University at Buffalo, Buffalo, New York, USA}
{shil@buffalo.edu}
{}
{}

\author{Juli\'{a}n Mestre}
{The University of Sydney, Sydney, Australia}
{julian.mestre@sydney.edu.au}
{}
{}

\author{Katina Russell}
{Georgetown University, Washington D.C., USA}
{katina.russell@cs.georgetown.edu}
{}
{}

\author{Seeun William Umboh}
{The University of Sydney, Sydney, Australia}
{william.umboh@sydney.edu.au}
{}
{}

\authorrunning{Cao et al.}

\Copyright{Nairen Cao, Jeremy T. Fineman, Shi Li, Juli\'{a}n Mestre, Katina Russell and Seeun William Umboh}



\relatedversion{https://doi.org/10.1145/3490148.3538554}

\begin{document}
\maketitle

\begin{abstract}
  The \defn{active-time scheduling problem} considers the problem of
  scheduling preemptible jobs with windows (release times and
  deadlines) on a parallel machine that can schedule up to $g$ jobs
  during each timestep.  The goal in the active-time problem is to
  minimize the number of active steps, i.e., timesteps in which at
  least one job is scheduled.  In this way, the active time models
  parallel scheduling when there is a fixed cost for turning the
  machine on at each discrete step.

  This paper presents a 9/5-approximation algorithm for a
  special case of the active-time scheduling problem in which job
  windows are laminar (nested). This result improves on the previous
  best 2-approximation for the general case.
\end{abstract}

\section{Introduction}
\label{section:intro}
The active-time scheduling is the problem of scheduling jobs with
windows on a parallel machine so as to minimize the number of 
timesteps during which machine is on, or active.

In the active-time problem~\cite{ChangGabow14}, we are given as input a set
$J$ of $n$ jobs, where each job $j\in J$ has an associated
processing time $p_j$, release time $r_j$, and deadline
$d_j\geq r_j+p_j$, all integers. The jobs are scheduled on a parallel
machine that can execute up to $g$ jobs during each step, where $g$ is
a positive integer specified as part of the input.  The input thus
comprises the jobs with their processing times $p_j$, release times
$r_j$, and deadlines $d_j$, as well as the machine parameter $g$, all
of which are integers.  Time is organized into discrete (integer)
steps or slots, and preemption is allowed but only at slot boundaries.
We call the time interval $[r_j,d_j)$ the job $j$'s \defn{window}.
Each job $j$ must be fully scheduled within its window, i.e., each job
must be assigned to $p_j$ timesteps, where each timestep $t$ to which
the job is assigned satisfied $r_j \leq t < d_j$.  Moreover, at most
$g$ jobs can be scheduled at any step.

We say that a timestep $t$ is \defn{active} if the schedule assigns at
least one job to step $t$.  The goal in the \defn{active-time
  scheduling problem} is to find a schedule with minimum number of
active steps that schedule all jobs within their windows.

We assume throughout that the instance is feasible.  Testing
feasibility (and producing a schedule) is an easy exercise applying
max flow.  In fact, this flow-based feasibility test generalizes to
any given subset of active timesteps (see, e.g., Appendix A.1
of~\cite{Kumar18}).  The active-time scheduling problem thus boils
down to figuring out which slots should be activated.

\subsection*{Problem History}

Chang, Gabow, and Khuller~\cite{ChangGabow14} introduce the active-time problem and show
that the problem can be solved optimally in polynomial time when the
processing times are all one. They also investigate various generalizations of the problem. For arbitrary integer processing
times, Chang, Khuller, and Mukherjee~\cite{Chang17} give two approximation algorithms. First, they give a rather 
complex rounding of the natural linear program (LP) that yields a 2-approximation.  They also show that the integrality gap of the natural LP is 2, so the rounding is tight. 
Second, they show that, any \defn{minimal feasible solution} yields a
3-approximation.  A minimal feasible solution is a set of active slots
such that (i) scheduling the jobs on those slots is feasible, and (ii)
deactivating any slot would render the schedule infeasible.  Consequently, a simple greedy algorithm (choose an
arbitrary active slot and deactivate it if the resulting set of slots
is still feasible) is a 3-approximation for the problem.  Kumar and
Khuller~\cite{Kumar18} give a greedy 2-approximation algorithm 
following the same general strategy of deactivating slots until reaching a minimal feasible solution, but they choose slots more carefully. They also exhibit inputs on which their algorithm achieves no better than a $2-1/g$ approximation, so the analysis is effectively tight. 

A key challenge for improving the approximation ratio for the general active-time problem is that the natural linear program has an integrality gap of at least $2-O(1/g)$, which converges to 2 as $g\rightarrow \infty$.  There is also no clear avenue to improve the 2-approximation obtained by Kumar and Khuller's~\cite{Kumar18} greedy approach. C\u{a}linescu and Wang~\cite{calinescu} suggest a stronger LP formulation that they conjecture has lower integrality gap, but they only show that the gap is at least $5/3$---whether it can lead to better approximations for general instances remains unknown.

\subsection*{Nested active-time scheduling  } To push past the barriers for the general version of the problem, this paper instead considers a
special case of the active-time problem in which the job windows are
laminar (nested).  That is, for each pair of jobs $i$, $j$, either the
intervals $[r_i,d_i)$ and $[r_j,d_j)$ are disjoint (meaning either
$d_i \leq r_j$ or $d_j \leq r_i)$), or one of the intervals is fully
contained in the other (i.e., either $r_i \leq r_j < d_j \leq d_i$ or
$r_j \leq r_i < d_i \leq d_j$). 

Our main result is a 9/5-approximation algorithm for the active-time problem with laminar job windows. Since the simple example exhibiting the integrality gap~\cite{Chang17} of 2 for the natural LP is a nested instance, a different LP formulation is needed.  Our algorithm starts by solving a stronger linear program (LP) for the problem
to produce a fractional solution, then performing a new rounding process over the tree of job windows.  The algorithm itself is not overly complex, but the analysis is not at all straightforward.  

Restricting our attention to nested windows gives us two advantages. 
First, assuming laminar windows allows us to augment the LP to obtain a smaller integrality gap than for the general case. Notably, 
our LP includes an additional ``ceiling constraint,'' which represents a stronger lower bound on the number of slots required in each window to the volume of jobs therein.  It is not clear how to take advantage of this same constraint in the general version of the problem.  As exhibited by our algorithm the integrality gap of our LP on the nested version of the problem is at most 5/3, which provides a separation between the nested and general versions of the problem. Secondly, the rounding process itself is inherently tied to the fact that nested windows form a tree. Even ignoring the issue of the larger integrality gap for the general case, it is not clear how to perform a similar rounding for general windows.

In \secref{gap}, we compare our strengthened LP formulation to that proposed by C\u{a}linescu and Wang's~\cite{calinescu} and show that both formulations exhibit an integrality gap of at least 3/2 for nested instances. In appendix, we show that the nested active time problem is NP complete.

\subsection*{Related work}

The objective of the active-time problem is motivated by an application to energy minimization. In this context, the machine can be turned off when no jobs are being executed and it takes the same amount of energy to run regardless of how many jobs are running---but it has a fixed capacity $g$ of how many jobs it can process on per active time slot. Energy-aware scheduling is an active area of research~\cite{albers,kirk} that is motivated by the pressing need of modern data centers whose large energy footprint accounts for most of their running costs~\cite{electricity-trends}.

There are many variations and generalization of the basic setup studied in this paper. Below we consider two of its most closely related variants. The reader is referred to the excellent survey by Chau and Li~\cite{chauli} for more related results such as online algorithms for active-time scheduling.

A natural generalization of the basic setup studied in this paper is to have, instead of a single interval, a collection of intervals where each each job can be scheduled. Chang \etal~\cite{ChangGabow14} show that this generalization is NP-hard when $g \geq 3$ even when jobs are unit-length, but that it can be solved in polynomial time when $g=2$. Furthermore, the problem admits an $H_g$-approximation for general $g$ via Wolsey's submodular cover framework~\cite{wolsey}.

Another related model is the busy-time problem where jobs cannot be preempted and we have parallel machines. This problem is much harder as even testing feasibility for a fixed number of machines is NP-hard. Indeed, the best approximation algorithm for minimizing the number of machines needed when $g=1$ is the $O\left(\sqrt{\frac{\log n}{\log \log n}}\right)$-approximation of Chuzhoy~\etal~\cite{chuzhoy}. Koehler and Khuller~\cite{koehler} show that it is possible minimize the number of machines use and simultaneously achieve a $O(1)$-approximation on the busy-time objective for instances with uniform processing time; for general instances with arbitrary processing times they can approximate the number of machines by a $O\left(\log \frac{p_{\max}}{p_{\min}}\right)$ factor while keeping the constant approximation bound for the busy-time objective.

\section{Preliminaries}
\label{section:preliminaries}
For an integer $p$, We use $[p]$ to represent integers from $\{1, 2, ..., p\}$.
Given an instance of the nested active time problem, we define its tree $T$ as follows. 
Each tree node $i$ is associated with an interval $K(i)$ such that $K(i) = [r_j, d_j)$ for some $j \in J$.  If there are several jobs with the same interval, we only create a single tree node. A tree node $i'$ is a child of $i$ if $K(i')\subsetneq K(i)$ and no other node interval is strictly between $K(i)$ and $K(i')$, i.e, there is no node $i''$ such that $K(i') \subsetneq K(i'') \subsetneq K(i)$. The descendants and ancestors of a node $i$ are denoted $\Des(i)$ and $\Anc(i)$, respectively.  Note that both $\Des(i)$ and $\Anc(i)$ include $i$ itself. To exclude $x$, we use $\Des^{+}(i) = \Des(i) \backslash \{ i \}$ and $\Anc^{+}(i) = \Anc(i) \backslash \{ i \}$. We define $\parent(i)$ to be the parent node of $i$. W.l.o.g we can assume $T$ is indeed a tree (instead of a forest) since otherwise the instance can be broken into several independent ones.



We assume that the tree contains $m$ nodes and each node is associated with an unique id in $[m]$. Now, each job $j$'s interval is associated with a node in the tree. For a job $j$, define $k(j)$ to be the tree node $i$ with $K( k(j) ) = [r_j, d_j)$; we say $j$ belongs to the node $i$ if $i = k(j)$. For jobs $j_1$ and $j_2$, if $r_{j_1} = r_{j_2}$ and $d_{j_1} = d_{j_2}$, then $k(i) = k(j)$.   Given a node $i$ and a job subset $J' \subseteq J$, $J'(i) = \{ j \in J' \mid k(j) = i \}$ is the set of jobs in $J'$ belonging to $i$.
Note that 
at least one job belongs to each node. 
Define the length of a node $i$, which is denoted as $L(i)$, as the $|K(i)| - \sum_{i' : \parent(i') = i}|K(i')|$, i.e, the number of time slots in the interval $K(i)$, but not in $K(i')$ for any child node $i'$ of $i$.


For simplicity, we use the following shorthand. Given a function or vector $f$ and a set $S$, if $f$ outputs reals, then $f(S) = \sum_{e \in S} f(e)$ or $f(S) = \sum_{e \in S}f_e$. If $f$ outputs subsets, then $f(S) = \union_{e \in S} f(e)$ or $f(S) = \union_{e \in S} f_e$.

We say that a node $i$ is \defn{rigid} if a feasible solution must open the entire interval $I(i)$.  For our rounding algorithm, it will be convenient if the tree is \defn{canonical}. 
\begin{definition}[Canonical trees]
A tree is \emph{canonical} if it is a binary tree and each leaf node is rigid.
\end{definition}

First, we transform an arbitrary tree to a binary tree.  If a parent node $i$ contains several children nodes $i_1, i_2, ..., i_t$, we will create several virtual nodes so that each node contains at most 2 children. Each virtual node's interval is the union of its children's intervals.  There are no jobs associated with the virtual nodes and the length of a virtual node $i'$ satisfying $L(i) = 0$.  Notice that this transformation adds at most $t$ virtual nodes for a node with $t$ children.  In total, this transformation only adds $k$ virtual nodes to a tree that had $k$ nodes originally.  In the resulting tree, only internal nodes can be virtual so each leaf node must have at least one job associated with it. 



We perform one final transformation to make each leaf node rigid. 
For a leaf node $i$, let $j \in J(i)$ be a job in $i$ with the longest processing time.  If $p_j = L(i)$, then we leave $i$ and the jobs therein unchanged.  Otherwise, we can assume that $j$ is scheduled in the first $p_j$ steps of $i$ because $j$ is the longest job in the leaf node and all jobs in the leaf node could choose the leaf's interval to fit in.  We transform the instance by creating a virtual child node $i'$ of the leaf $i$ with interval corresponding to the first $p_j$ steps of $I(i)$, and we reduce $j$'s window to match $i'$'s.  
Notice this transformation does not change our solution for the original tree.

\section{Algorithm}
\subsection{Linear Program}
\label{subsection:linearprogram}

\begin{figure}
  \begin{subfigure}[c]{.58\linewidth}
        \begin{equation}
        \textbf{min} \quad \sum_{i \in [m]} x(i) \quad \textbf{s.t.}  \label{LP}
    \end{equation}\vspace*{-10pt}
    \begin{align}
        \sum_{i \in \Des(K(j)) } y(i, j) \ge p_j, \quad&\forall j \label{LPC:scheduled} \\
        \sum_{j \in J(Anc(i))} y(i, j) \le g \cdot x(i), \quad&\forall i \label{LPC:capacity-g} \\
        x(i) \le L(i), \quad&\forall i \label{LPC:capacity-L} \\
        y(i, j) \le x(i), \quad&\forall i, j \label{LPC:y-less-than-x} \\
        y(i, j) = 0, \quad&\forall i, j \notin J(Anc(i)) \label{LPC:y-equal-0}  \\
        \sum_{i' \in \Des(i)} x(i') \ge 2, \quad&\forall i, \text{OPT}_i \geq 2 \label{LPC:strong1}
        \\
        \sum_{i' \in \Des(i)} x(i') \ge 3, \quad&\forall i, \text{OPT}_i \geq 3 \label{LPC:strong2}
    \end{align}
\caption{Linear program for active time scheduling.  
    By default we restrict $i \in [m]$ and $j \in J$.}
\label{figure:lp}
  \end{subfigure}\hfill
  \begin{tabular}[c]{@{}c@{}}
    \begin{subfigure}[c]{.38\linewidth}
      \centering
      \includegraphics[width=\linewidth]{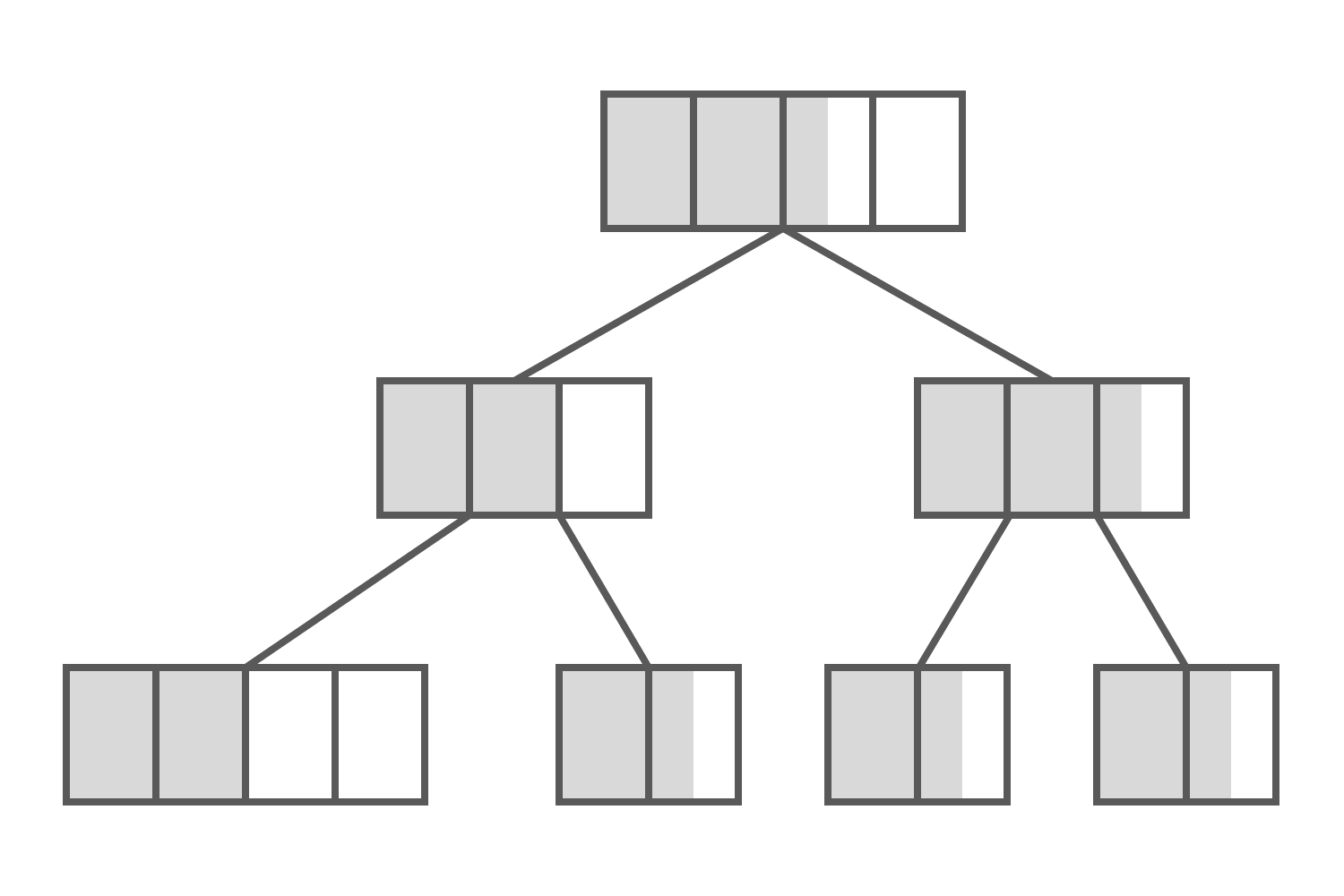}%
      \caption
        {%
          The open slots from an LP solution
          \label{figure:LPtransformation1}%
        }%
    \end{subfigure}\\
    \noalign{\bigskip}%
    \begin{subfigure}[c]{.38\linewidth}
      \centering
        \includegraphics[width=\linewidth]{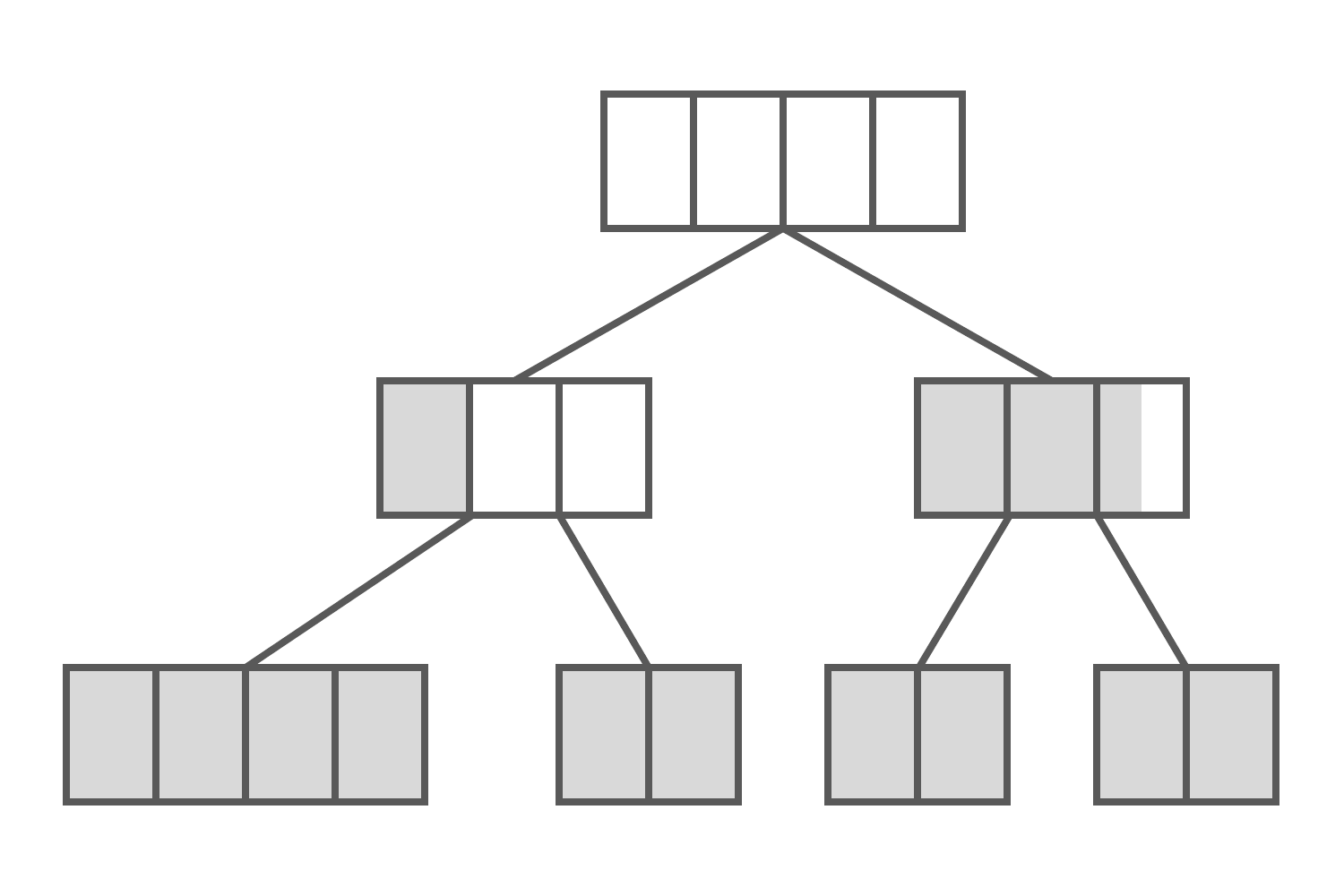}
      \caption
        {%
          The open slots after performing the LP transformation
          \label{figure:LPtransformation2}%
        }%
    \end{subfigure}
  \end{tabular}
    \caption{(a) is the linear program for active time schuduling. (b) and (c) are an example of a tree before and after running the LP transformation in Lemma \ref{lemma:LPtransform}. The dark slots represent slots have jobs scheduled in them, and the white slots are closed.}
    \label{fig:lpmain}
\end{figure}

The linear program is LP \eqref{LP}, which is given in Figure \ref{fig:lpmain}(a). In the LP, $x(i)$ denotes the number of time slots opened in node $i$, and $y(i,j)$ denotes the amount of job $j$  that is scheduled in node $i$.   In the LP, $\text{OPT}_i$ denotes the smallest number of slots to schedule the jobs in $J(\Des(i))$.





The objective is to minimize $\sum_{i \in [m]} x(i)$.
\eqref{LPC:scheduled} ensures that every job $j$ is scheduled in at least $p_j$ time slots.  \eqref{LPC:capacity-g} ensures that the total number of jobs scheduled in $x(i)$ is at most $g \cdot x(i)$, for each node $i \in [m]$. \eqref{LPC:capacity-L} requires that the number of open time slots in a node $x(i)$ is at most the interval length $L(i)$ of node $i$. \eqref{LPC:y-less-than-x} says that we could give at most $x(i)$ time slots for a job $j$. 
\eqref{LPC:y-equal-0} restricts that for each job $j \in J$, $j$ can only be put into nodes in  $\Des(K(j))$.
%
%
\eqref{LPC:strong1} and \eqref{LPC:strong2}
are the key constraints that makes the LP stronger. They are clearly valid; moreover, checking if $\text{OPT}_i \geq 2$ ($\text{OPT}_i \geq 3$) can be done easily.

For simplicity, given a node set $V \subset [m]$ and $J' \subset J$, $y(V, J') = \sum_{i \in V, j \in J'} y(i, j)$; If either $V$ or $J'$ is a singleton, we can replace it by the unique element in it.  Let $x(S) = \sum_{i \in S} x(i)$ for every $S \subseteq [m]$.

After running the LP and getting a solution $(x, y)$, we will perform a transformation on the solution.

\subsection{Transformation of LP Solution}
\begin{lemma} \label{lemma:LPtransform}
Given a feasible LP solution, we can efficiently output another feasible LP solution such that for any pair of nodes $i_1$, $i_2$ such that $i_2 \in \Des^+(i_1)$, if $x(i_2) < L(i_2)$, then $x(i_1) = 0$. 
\end{lemma}
\begin{proof}
    Suppose there are nodes $i_1, i_2$ with $i_2 \in \Des^+(i_1)$ and $x(i_2) < L(i_2)$ and $x(i_1) > 0$.  Then let $\theta = \min\{L(i_2) - x(i_2), x(i_1)\} > 0$. We can move $\theta$ fractional open slots from $i_1$ to $i_2$. For every job $j$, we move $\frac{\theta}{x(i_1)} y(i_1, j)$ fractional assignment of $j$ from $i_1$ to $i_2$. More specifically, let $(x', y') = (x, y)$ initially. We decrease $x'(i_1)$ by $\theta$ and increase $x'(i_2)$ by $\theta$.  For every $j \in J$, we decrease $y'(i_1, j)$ to $\frac{x'(i_1)}{x(i_1)} y(i_1, j)$ and increase $y'(i_2, j)$ to $y'(i_2, j) + \frac{\theta}{x(i_1)}y(i_1, j)$. Notice that every job $j$ that can be assigned to $i_1$ can also be assigned to $i_2$. It is not hard to show that all constraints remain satisfied by the new solution $(x', y')$. Finally we update $(x, y) \gets (x', y')$.
    
    Notice that after the operation, we have either $x(i_1) = 0$ or $x(i_2) = L(i_2)$. By repeating the procedure polynomial number of times, we can find a solution $(x, y)$ satisfying the property of the lemma.
\end{proof}

An example of of the LP transformation is shown in Figure \ref{fig:lpmain}(b) and \ref{fig:lpmain}(c).
Lemma \ref{lemma:LPtransform} implies that for any $i$ with $x(i)>0$, all of its strict descendants are fully open. We let $I$ be the set of topmost nodes $i$ with $x(i) > 0$; those are the nodes $i$ with $x(i) > 0$ but all its strict ancestors $i'$ have $x(i') = 0$. 
\begin{claim}
	\label{claim:properties-of-I}
	The following properties hold for $I$:
	\begin{enumerate}[label=(\ref{claim:properties-of-I}\alph*)]
	    \item No node in $I$ is a strict ancestor of another node in $I$.
	    \item $\Des(I)$ contains all leaves. 
	    \item Every $i \in I$ has $x(i) > 0$.
	    \item For any $i \in I$ and $i' \in \Des^+(i)$, we have $x(i') = L(i')$.
	    \item For any $i \in I$ and $i' \in \Anc^+(i)$, we have $x(i') = 0$. 
	\end{enumerate}
\end{claim}

	We make one modification to the tree that do not change the instance. For every $i \in \Anc^+(I)$, we can assume that $i$ has exactly two children: if $i$ has only one child, we remove it from the tree and connect its parent directly to its children. This does not change the instance since $x(i) = 0$.

	We also want to mention for any node $i$ such that $x(\Des(i)) \in (1, 2)$, $i$ has one child $i'$, which is a leaf with $x(i') = L(i') = 1$ because of the rigidity of the leaf node.

\subsection{Rounding Algorithm to Obtain an Integral Vector $\tilde x \in \{0, 1\}^{[m]}$}	

	The rounding algorithm that constructs our integral $\tilde x$ is given in Algorithm~\ref{alg:rounding}.
	\begin{algorithm}[H]
		\caption{Rounding Algorithm}
		\label{alg:rounding}
		\begin{algorithmic}[1]
			\State let $\tilde x(i) \gets \floor{x(i)}, \forall i \in I$ and $\tilde x(i) \gets x(i), \forall i \in [m] \setminus I$. 
			\For{every node $i \in \Anc(I)$ from bottom to top}
				\While{$\frac{9x(\Des(i))}{5} \geq \tilde x(\Des(i)) + 1$}
					\If{$\exists i' \in \Des(i)$ with $\tilde x(i') <  x(i')$}
						\State choose such an $i'$ arbitrarily
						\State let $\tilde x(i') \gets \ceil{x(i')}$
					\Else
						\State break
					\EndIf
				\EndWhile
			\EndFor
		\end{algorithmic}
	\end{algorithm}
	
	Clearly, the number of open slots is at most $\frac{9 x([m])}{5}$.
	
\begin{lemma}
\label{lemma:ratio}
After running the Algorithm \ref{alg:rounding}, $\tilde x([m]) \leq \frac{9 x([m])}{5}$.
\end{lemma}

\section{Feasibility of $\tilde x$}
\label{section:analysis}
In this section, we show that the rounded time slot $\tilde x$ is a feasible solution.  

\subsection{A Necessary and Sufficient Condition}
First we will give an if-and-only-if condition. From now on, for any set $J' \subseteq J$, we define $J'(i) = J(i) \cap J'$ for every $i \in [m]$ and $J'(S) = J(S) \cap J'$ for every $S \subseteq [m]$. 
\begin{lemma}
\label{ifandonlyif}
    Given an integer solution $\tilde x$ for the LP, $\tilde x$ is feasible if and only for every subset $J' \subseteq J$ of jobs, we have 
	\begin{align}
		\sum_{i \in [m]} \min\{|J'(\Anc(i))|, g\} \cdot \tilde x(i) \geq p(J'). \label{inequ:if-and-only-if}
	\end{align}
\end{lemma}
\begin{proof}
The only if part is easy to see. Any node $i$ can hold at most $\min\{|J'(\Anc(i))|, g\} \cdot \tilde x(i)$ volume of jobs in $J'$. All the jobs in $J'$ should be assigned. If $\tilde x$ is feasible, then $\sum_{i \in [m]} \min\{|J'(\Anc(i))|, g\} \cdot \tilde x(i) \geq p(J')$.

Now we prove the if part, by considering the contra-positive of the statement and applying the maximum-flow-minimum cut theorem. Assume $\tilde x$ is not feasible. 
We construct a 4-layer directed network $H = (V_H, E_H)$, where the nodes from left to right are $\{s\}, J, [m] $ and $\{t\}$. There is an edge from $s$ to every $j \in J$ with capacity $p_j$, an edge from every $j \in J$ to every $i \in \Des(k(j))$ with capacity $\tilde x(i)$, and an edge from every $i \in [m]$ to $t$ with capacity $g\cdot \tilde x(i)$. For a subsets $V'\subseteq V_H$ and a node $v \in V_H \setminus V'$, we use $E_H(V', v)$ to denote the set of edges from $V'$ to $v$. 

As $\tilde x$ is not feasible, there is a $s$-$t$ cut in the network with capacity less than $p(J)$. Let $(A, B)$ be the cut: $s \in A, t \in B$ and $A \uplus B = V_H$ . Its cut value, which is
$p(B \cap J) + \sum_{i \in B \cap [m]}|E_H(A \cap J, i)|\cdot \tilde x(i) + g\cdot \tilde x(A \cap [m])$, is less than $p(J)$. This is equivalent to 
$\sum_{i \in B \cap [m]} |E_H(A \cap J, i)|\cdot \tilde x(i) + g\cdot \tilde x(A \cap [m]) < p(A \cap J)$. Let $J' = A \cap J$. Then the contribution of a node $i \in [m]$ to the left-side is either $|E_H(J', i)|\cdot\tilde x(i)$ (if $i \in B$), or $g \cdot \tilde x(i)$ (if $i \in A$), which is lower bounded by $\min\{|E_H(J', i)|, g\}\cdot \tilde x(i)$. Noticing $|E_H(J', i)| = |J'(\Anc(i))|$, we have $\sum_{i \in [m]} \min\{|J'(\Anc(i))|, g\}\cdot \tilde x(i) < p(J')$. This finishes the proof of the if part. 
\end{proof}

\begin{lemma}
	\label{lemma:J'-irreducible}
	In Lemma~\ref{ifandonlyif}, it is sufficient to consider the sets $J' \subseteq J$ satisfying the following property: 
	\begin{enumerate}[label=(\ref{lemma:J'-irreducible}\alph*)]
		\item $p_j > \tilde x \Big(\big\{i \in \Des(K(j)): |J'(\Anc(i))| \leq g\big\}\Big), \forall j \in J'$.
	\end{enumerate}
\end{lemma}
\begin{proof}
	Suppose $J'$ does not satisfy the property. Then for some $j \in J'$ we have  $\tilde x \Big(\big\{i \in \Des(K(j)): |J'(\Anc(i))| \leq g\big\}\Big) \geq p_j$.  Then removing $j$ from $J'$ will  decrease the left side of \eqref{inequ:if-and-only-if} by $\tilde x \Big(\big\{i \in \Des(K(j)): |J'(\Anc(i))| \leq g\big\}\Big)$, and the right side by $p_j$.  Thus, the inequality \eqref{inequ:if-and-only-if} for $J'$ will be implied by the inequality for $J' \setminus \{j\}$.
\end{proof}

Once we have the if-and-only-if condition for the feasibility, the main lemma we need to prove is the following:
	\begin{theorem}
		\label{thm:covering}
		For every subset $J' \subseteq J$ of jobs satisfying the property in Lemma \ref{lemma:J'-irreducible}, we have \eqref{inequ:if-and-only-if}.
	\end{theorem}
	
	 The rest of the section is dedicated to the proof of Theorem~\ref{thm:covering}. From now on we fix a subset $J' \subseteq J$ satisfying the property of Lemma~\ref{lemma:J'-irreducible}. Our goal is to prove \eqref{inequ:if-and-only-if}. 
	 
	 For notational convenience, let  $u_i = \min\{|J'(\Anc(i))|, g\}$ and $w_i = u_i x(i)$ and $\tilde w_i = u_i \tilde x(i)$ for every $i \in [m]$. Thus, \eqref{inequ:if-and-only-if} is simply written as $p(J') \leq \tilde w([m])$. Recall that $y(V, J') = \sum_{i \in V, j \in J'} y(i, j)$ for a given $V \subseteq [m], J' \subseteq J$. We have $p(J') = y([m], J') = y(\Des(I), J')$ and $\tilde w([m]) = \tilde w(\Des(I))$. Thus, we need to prove 
	 \begin{align}
	 	y(\Des(I), J') \leq \tilde w(\Des(I)), \label{inequ:reformulate}
	 \end{align}
	 for every $J' \subseteq J$ satisfying the property in Lemma~\ref{lemma:J'-irreducible}.
	 
	 The following simple claim will be used multiple times:
	\begin{claim}
		\label{claim:simple}
		For every $i \in [m]$, we have $y(i, J') \leq w_i = u_i x(i)$. 
	\end{claim}
	\begin{proof}
		If $u_i = g$, we use \eqref{LPC:capacity-g} in the LP. If $u_i < g$, then we use \eqref{LPC:y-less-than-x} and \eqref{LPC:y-equal-0}.
	\end{proof}

\subsection{Construction of Triples}

	For nodes $i \in [m] \setminus I$, we have $\tilde x(i) = x(i)$.  For nodes $i \in I$ with $x(\Des(i)) \notin (1, \frac{10}{9})$, we have $\tilde x(i) = \ceil{x(i)}$ since $\frac{5 x(\Des(i))}{3} \geq \ceil{x(\Des(i))}$.  Thus, for these nodes $i$, Claim~\ref{claim:simple} implies $y(i, J') \leq \tilde w_i$.   The critical nodes are those $i \in I$ with $x(\Des(i)) \in (1, \frac{10}{9})$.
	
	With this in mind, we classify nodes in $I$ into two types: a node $i \in I$ is of 
	\begin{itemize}
		\item type-$\sfB$ if $x(\Des(i)) \in \{1\} \cup [\frac43, \infty)$, and
		\item type-$\sfC$ if $x(\Des(i)) \in (1, \frac43)$.
	\end{itemize}
	In the definition, we use $\frac43$ instead of $\frac{10}{9}$ to create some buffers.  Furthermore, a type-$\sfC$ node $i \in I$ is of 
	\begin{itemize}
		\item type-$\sfC_1$ if $\tilde x(\Des(i)) = 1$, and 
		\item type-$\sfC_2$ if $\tilde x(\Des(i)) = 2$.
	\end{itemize}
    
    \textbf{At most 2 type-$\sfC$ nodes} Before going to the triples, we first solve the case at most 2 type-$\sfC$ nodes are in $I$. At the same time, if 1 type-$\sfB$ node exists, then all type-$\sfC$ nodes are type-$\sfC_2$. 
    \begin{lemma}
	\label{lemma:at-most-2-typec}
	If there are at most 2 type-$\sfC$ nodes and at least 1 type-$\sfB$ node in $I$, then all type-$\sfC$ are type-$\sfC_2$.
    \end{lemma}	
    \begin{proof}
    Let $i_1$ and $i_2$(if exists) be the type-$\sfC$ node and $i_3$ be the type-$\sfB$ node. Notice that $\frac{9}{5} x(i_3) - \ceil{x(i_3)} \geq 0.4$. We have $ \frac{9}{5}(x(i_1) + x(i_3)) \geq x(i_1) + 0.8 + \ceil{x(i_3)} + 0.4 \geq \ceil{x(i_1)} + \ceil{x(i_3)}$ and $\frac{9}{5}(x(i_1) + x(i_2) + x(i_3)) \geq x(i_1) + 0.8 + x(i_2) + 0.8 + \ceil{x(i_3)} + 0.4 \geq \ceil{x(i_1)} + \ceil{x(i_2)}+ \ceil{x(i_3)}$. In either case, algorithm~\ref{alg:rounding} can afford to round up all type-$\sfC$ nodes.
    \end{proof}
    
    Based on Lemma~\ref{lemma:at-most-2-typec}, if at least one type-$\sfB$ node is in $I$, then we already rounded up all type-$\sfC$ nodes. Assume that there  no type-$\sfB$ node in $I$, then we have $x([m]) \leq 2 \times 4/3 = 8/3$. due to the LP constraint \eqref{LPC:strong1} and \eqref{LPC:strong2}, $x(i)$ are integer for all $i \in [m]$ and this contradicts to the assumption that no type-$\sfB$ node is in $I$.
	
	
	\textbf{More than 2 type-$\sfC$ nodes} When we have at least 3 type-$\sfC$ nodes in $I$, we want to create disjoint triples of type-$\sfC$ nodes. Each triple contains 1 type-$\sfC_1$ node, and 2 type-$\sfC_2$ nodes. Moreover, all the type-$\sfC_1$ nodes are contained in these triples. Later for each triple $(i_1, i_2, i_3)$ we constructed, we shall prove $y(\Des(\{i_1, i_2, i_3\}), J') \leq \tilde w(\Des(\{i_1, i_2, i_3\}))$. This will prove \eqref{inequ:reformulate}.

	The construction of triples are given in Algorithm~\ref{alg:cnstr-triples}. Notice that this is not a part of our algorithm for solving the active time scheduling problem; it is only used in the analysis.   If we have a type-$\sfC_1$ node $i_1$ and a type-$\sfC_2$ node $i_2$ as brothers, then we say $(i_1, i_2)$ is a $\sfC_1\sfC_2$-brother-pair.  In our triples, we make sure that we do not break $\sfC_1\sfC_2$-brother-pairs: for such a pair $(i_1, i_2)$, there must be some $\sfC_2$-node $i_3$ such that $(i_1, i_2, i_3)$ is a triple we constructed.

\begin{algorithm}[h]
	\caption{Construction of Triples}
	\label{alg:cnstr-triples}
	\begin{algorithmic}[1]
		\State $\triples \gets \emptyset$, set all type-$\sfC_1$ nodes as uncovered, and all type-$\sfC_2$ nodes as unused.
		\For{every node $i \in \Anc(I)$ with $|\Des(i) \cap I| \geq 3$ from bottom to top}
			\While{$\exists$ an uncovered type-$\sfC_1$ node $i_1 \in \Des(i)$}
				\State \label{step:choose-C2-nodes} choose two unused type-$\sfC_2$ nodes $i_2, i_3 \in \Des(i)$, without breaking $\sfC_1\sfC_2$-brother-pairs 
				\State add $(i_1, i_2, i_3)$ to $\triples$, claim $i_1$ is covered, and $i_2$ and $i_3$ are used.
			\EndWhile
		\EndFor
	\end{algorithmic}
\end{algorithm}

\begin{lemma}
	\label{lemma:can-cover}
	In Step~\ref{step:choose-C2-nodes} of Algorithm~\ref{alg:cnstr-triples}, there are at least two unused type-$\sfC_2$ nodes in $\Des(i)$.
\end{lemma}	
\begin{proof}
	Let $n_1$ and $n_2$ be the number of type-$\sfC_1$ and type-$\sfC_2$ nodes in $\Des(i) \cap I$ respectively.  Let $n' = n_1 + n_2$.  We shall prove $n_2 \geq 2n_1$, which implies that we will not run out of type-$\sfC_2$ nodes and proves the lemma. 
		
	First, we consider the case where there is no type-$\sfB$ node in $\Des(i) \cap I$.  Then $n' = |\Des(i) \cap I| \geq 3$. Moreover, $n_1 + 2n_2 \geq \floor{\frac{9n'}{5}}$ or $n_1 = 0$ by our rounding algorithm in Algorithm~\ref{alg:rounding}. In the latter case we clearly have $n_2 \geq 2n_1$. So, we assume the former.  Then $n_2 \geq \floor{\frac{4n'}{5}}$.  Notice that $\floor{\frac{4n'}{5}} \geq \frac{4n'}{5} - \frac{4}{5}$ and thus $\floor{\frac{4n'}{5}} \geq \frac{2n'}{3}$ whenever $n' \geq 6$.  One can check that when $n' \in \{3, 4, 5\}$ we have $\floor{\frac{4n'}{5}} \geq \frac{2n'}{3}$. Therefore if $n' \geq 3$, we have $\floor{\frac{4n'}{5}} \geq \frac{2n'}{3}$. So $n_2 \geq 2n_1$.
	
	Now we consider the other case where there is at least one type-$\sfB$ node in $\Des(i) \cap I$.  If $n_1 = 0$, then $n_2 \geq 2n_1$ and thus we assume $n_1 > 0$.  Then we have $n_1 + 2n_2 \geq \floor{\frac{9n'}{5} + \frac25}$, as the type $\sfB$ node $i^*$ have $\frac{9 x(\Des(i^*))}{5} \geq \ceil{x(\Des(i^*))} + \frac25$. This implies $n_2 \geq \floor{\frac{4n'}{5} + \frac25}$.  Then $n_2 \geq \frac{2n'}{3}$ as $\floor{\frac{4n'}{5} + \frac25} \geq \frac{2n'}{3}$ for every integer $n' \geq 0$. Thus, $n_2 \geq 2n_1$. 
\end{proof}

\begin{figure}
	\includegraphics[width=0.8\linewidth]{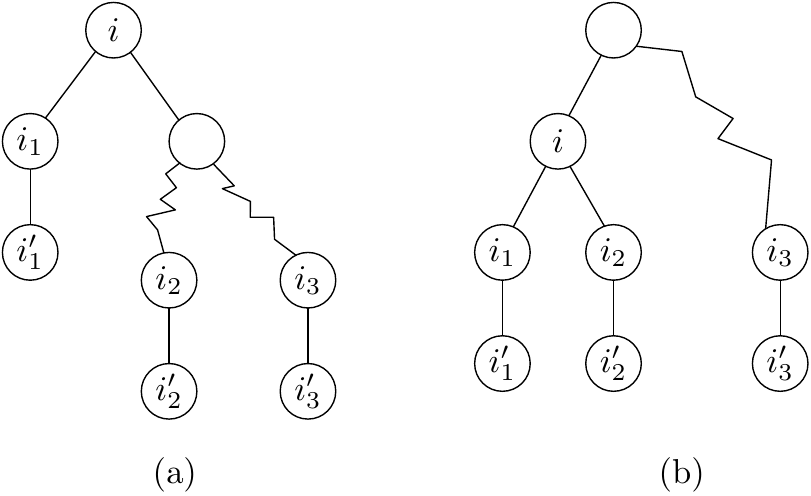}
	\caption{The two cases in Lemma~\ref{lemma:two-cases} and \ref{lemma:bound-each-triple}.}
	\label{fig:twocases}.
\end{figure}

\subsection{Proof of \eqref{inequ:reformulate} Using the Triples}
In this section, we prove \eqref{inequ:reformulate} by using the constructed triples. First, we show that they satisfy some good properties: 

\begin{lemma}
	\label{lemma:two-cases}
	For every $(i_1, i_2, i_3) \in \triples$, one of the following two conditions hold:
	\begin{enumerate}[label=(\ref{lemma:two-cases}\alph*)]
		\item \label{item:case1} $i_2, i_3 \in \Des^+(\parent(i_1))$. 
		\item \label{item:case2} $i_1$ and $i_2$ are brothers, and $i_3 \in \Des^+(\parent(\parent(i_1)))$.
	\end{enumerate} 
\end{lemma}

\begin{proof}
	Consider any type-$\sfC_1$ node $i_1$. Let $i^*$ and $i'$ be the parent and brother of $i_1$ respectively. 

	First consider the case that $i' \notin I$. By (\ref{claim:properties-of-I}a) and (\ref{claim:properties-of-I}b), we have  $|\Des(i^*) \cap I| \geq 3$.  By Algorithm~\ref{alg:cnstr-triples} and Lemma~\ref{lemma:can-cover}, $i'$ will be covered when we are at the iteration $i = i^*$ in Algorithm~\ref{alg:cnstr-triples}. So, the first property holds.
	
	Then consider the other case that $i' \in I$. Then $i'$ can not be of type-$\sfB$ since otherwise $i_1$ should be of type-$\sfC_2$. $i'$ can not be of type-$\sfC_1$ since we could open $3$ slots in $\Des(i^*)$. Therefore $i'$ must be of type-$\sfC_1$ and $(i_1, i')$ is a $\sfC_1\sfC_2$-brother-pair. In this case the second property holds.
\end{proof}

See Figure~\ref{fig:twocases} for the two cases obtained in Lemma~\ref{lemma:two-cases}, which will be used again in the proof of the following Lemma:
\begin{lemma}
	\label{lemma:bound-each-triple}
	For every $(i_1, i_2, i_3) \in \triples$, we have 
	\begin{align*}
		y(\Des(\{i_1, i_2, i_3\}), J') \leq \tilde w(\Des(\{i_1, i_2, i_3\})).
	\end{align*}
\end{lemma}


\begin{proof}
	Recall that $i_1$ is of type $\sfC_1$ and $i_2$ and $i_3$ are of type-$\sfC_2$. 
	Let  $i'_1, i'_2$ and $i'_3$ be the children of $i_1, i_2$ and $i_3$ respectively.  By Lemma~\ref{lemma:two-cases},  either \ref{item:case1} or \ref{item:case2} holds.
	
	We first assume \ref{item:case1}. Let $i = \parent(i_1)$.  See Figure~\ref{fig:twocases}(a) for all illustration of nodes used in this case. 
	
	By \eqref{LPC:strong1} in the LP, $J(\Des(i_1))$ can be scheduled in the one slot in $i'_1$ since $x(\Des(i_1)) < 2$. That is, all the jobs in the set has size 1 and there are at most $g$ of them.  So, we have 
	\begin{align}
		& y(\Des(i_1), J'(\Des(i_1))) \leq |J'(\Des(i_1))| \leq u_{i'_1} x(i'_1).  \label{inequ:case1-1}\\
		& y(\Des(i_1), J'(\Anc^+(i_1))) \leq \frac{10}{9}\min\{|J'(\Anc^+(i_1))|, g\} \nonumber \\
		&\qquad \leq (\tilde x(i_2) - x(i_2) + \tilde x(i_3) - x(i_3))\min\{|J'(\Anc^+(i_1))|, g\}\nonumber\\
		&\qquad \leq (\tilde x(i_2) - x(i_2)) u_{i_2} + (\tilde x(i_3) - x(i_3)) u_{i_3} \label{inequ:case1-2} \\
		& y(\Des(\{i_2, i_3\}), J') \leq u_{i_2} x(i_2) + u_{i'_2}  x(i'_2) + u_{i_3} x(i_3) + u_{i'_3} x(i'_3) \label{inequ:case1-3}
	\end{align}
	The first inequality of \eqref{inequ:case1-1} is by that all jobs in $J'(\Des(i_1))$ have size $1$,  and the second one is by $u_{i'_1} = \min\{|J'(\Anc(i'_1))| ,g\} \geq |J'(\Des(i_1))|$ and $x(i'_1) = 1$. The first inequality of \eqref{inequ:case1-2} is by that $x(\Des(i_1)) < \frac{10}{9}$. The second one follows from $\tilde x(i_2) - x(i_2) \geq \frac{2}{3}$ and $\tilde x(i_3) - x(i_3) \geq \frac{2}{3}$. The last one used that every job in $J'(\Anc^+(i_1))$ can be assigned to $i_2$ and $i_3$.  \eqref{inequ:case1-3} is by Claim~\ref{claim:simple}.
	
	Adding \eqref{inequ:case1-1}, \eqref{inequ:case1-2} and \eqref{inequ:case1-3}, we have
	\begin{align*}
		\quad y(\Des(\{i_1, i_2, i_3\}), J') &\leq u_{i'_1} x(i'_1) + u_{i_2}\tilde x(i_2) + u_{i'_2} x(i'_2) + u_{i_3}\tilde x(i_3) + u_{i'_3} x(i'_3)\\
		&= \tilde w(\Des(\{i_1, i_2, i_3\}))
	\end{align*}

	Now we consider the case that \ref{item:case2} holds.  Let $i = \parent(i_1) = \parent(i_2)$.  See Figure~\ref{fig:twocases}(b) for illustration of the nodes.

	First, if $u_{i_2} = g$, then we have $(\tilde x(i_2) - x(i_2)) u_{i_2} \geq (x(i_1) - \tilde x(i_1)) u_{i_1}$ as $\tilde x(i_2) - x(i_2) \geq \frac23$ and $x(i_1) - \tilde x(i_1) < \frac19$. This is  $\tilde w(i_1) + \tilde w(i_2) \geq w(i_1) + w(i_2)$. Thus,  $y(\Des(\{i_1, i_2, i_3\}), J') \leq w(\Des(\{i_1, i_2, i_3\})) \leq \tilde w(\Des(\{i_1, i_2, i_3\}))$ as $\forall i' \in \{i_3, i'_1, i'_2, i'_3\}$ we have $w(i') \leq \tilde w(i')$.
	
	So assume $u_{i_2} < g$.  As we assumed $J'$ satisfies (\ref{lemma:J'-irreducible}a), every job $j \in J'(\Anc(i_2))$ has $p_j > 1$.  All jobs in $J(i)$ have $p_j \leq 2$ since otherwise we would have $x(\Des(i)) \geq 3$ by LP constraint \eqref{LPC:scheduled} and \eqref{LPC:y-less-than-x}. Also all jobs in $J(i_2)$ have size 1. So all jobs in $J'(i)$ have size $2$, and $J'(i_2) = \emptyset$.
	
	Then we have	
	\begin{align}
		& y(\Des(i_1), J'(\Des(i_1))) + |J'(i)| \leq |J'(\Des(i_1))| + |J'(i)| \leq u_{i'_1} x(i'_1). \label{inequ:case2-1} \\
		& y(\Des(i_1), J'(\Anc^+(i))) \leq \frac{10}{9} |J'(\Anc^+(i))| \leq \big(\tilde x(i_2) - x(i_2) + \tilde x(i_3) - x(i_3)\big) |J'(\Anc^+(i))| \nonumber\\
		&\qquad\leq (\tilde x(i_2) - x(i_2)) |J'(\Anc^+(i))| + (\tilde x(i_3) - x(i_3)) u_{i_3} \label{inequ:case2-2}\\
		& y(\Des(\{i_2, i_3\}), J'\setminus J'(i)) + |J'(i)|  \leq |J'(\Anc^+(i))|\cdot x(i_2) + u_{i'_2}  x(i'_2) + u_{i_3} x(i_3) \nonumber\\
		&\qquad \quad + u_{i'_3} x(i'_3) + |J'(i)|\cdot \tilde x(i_2) 
		\label{inequ:case2-3}
	\end{align}
	
	The first inequality of \eqref{inequ:case2-1} all jobs in $J'(\Des(i_1))$ have size 1, and the second inequality is by that $|J'(\Des(i_1))| + |J'(i)| \leq g$ since otherwise $\text{OPT}_{i} \geq 3$.  The proof of \eqref{inequ:case2-2} is similar to that of \eqref{inequ:case1-2}. Notice that every job in $J'(\Anc^+(i))$ can be assigned to $i_2$ and $i_3$.  The inequality in \eqref{inequ:case2-3} used Claim~\ref{claim:simple} for $i'_2, i_3$ and $i'_3$.

	Adding \eqref{inequ:case2-1}, \eqref{inequ:case2-2} and \eqref{inequ:case2-3}, we get 
	\begin{align*}
		&\quad y(\Des(i_1, i_2, i_3), J' \setminus J'(i)) + 2|J'(i)| \\
		&\leq u_{i'_1}x(i'_1) + (|J'(\Anc^+(i)| + |J'(i)|) \tilde x(i_2) + u_{i'_2} x(i'_2) + u_{i_3} \tilde x(i_3) + u_{i'_3}x(i'_3)\\
		&=u_{i'_1}x(i'_1) + u_{i_2} \tilde x(i_2) + u_{i'_2} x(i'_2) + u_{i_3} \tilde x(i_3) + u_{i'_3}x(i'_3)= \tilde w(\Des(i_1, i_2, i_3)).
	\end{align*}
	Notice that $y(\Des(i_1, i_2, i_3), J'(i)) = p(J'(i)) = 2|J'(i)|$, we have $y(\Des(i_1, i_2, i_3), J') \leq \tilde w(\Des(i_1, i_2, i_3))$.
\end{proof}

With Lemma~\ref{lemma:bound-each-triple}, we can prove \eqref{inequ:reformulate}. First $\sum_{i \in *} y(\Des(i), J') \leq \sum_{i \in *} \tilde w(\Des(i))$, where $i \in *$ is over all nodes in the triples we constructed. For all the other nodes $i \in I$, we have $y(\Des(i), J') \leq w(\Des(i)) \leq \tilde w(\Des(i))$.  Therefore we have $y(\Des(I), J') \leq \tilde w(\Des(I))$, which is exactly \eqref{inequ:reformulate}. Combining with Lemma~\ref{lemma:ratio}, we have following theorem,

\begin{theorem}
\label{1.8main}
There exists a 1.8-approximation polynomial-time algorithm for the nested active-time problem.
\end{theorem}
	

\section{Integrality gap}
\label{sec:gap}

For the general (non-nested) version of the active scheduling problem, C\u{a}linescu and Wang~\cite{calinescu} proposed a slightly stronger than our LP from Figure~\ref{figure:lp} and showed a non-nested instance where the integrality gap approaches $5/3$ as $g \rightarrow \infty$. In this section, we show that their LP and our LP have an integrality gap of at least $3/2$ on nested instances. 

To define their LP, we need some notation. Let $\mathcal{T} = [\min_{j \in J} r_j, \\ \max_{j \in J} d_j )$ denote the set of time steps between the earliest release time and the latest deadline. For an interval of time $I = [t_1, t_2)$ for some $t_1, t_2 \in \mathcal{T}$ and a job $j$, let $q_j(I)$ denote the minimum number of slots within $I$ that job $j$ needs to occupy in a feasible solution even if all slots outside of $I$ were active and available to $j$. The variable $x(t)$ denotes the extent to which the slot $t$ is active and $y(t,j)$ denotes the extent to which job $j$ is assigned to slot $t$. See Figure~\ref{figure:calinescu-lp}.

\begin{figure}
    \begin{equation*}
    \boxed{
    \begin{aligned}
    \textbf{min} \sum_{t \in \mathcal{T}} x(t) \\
    \textbf{s.t.} 
    \sum_{t \in [r_j, d_j)} y(t,j) \geq p_j \qquad & \forall j \in J\\
    \sum_{j \in J} y(t,j) \leq g\cdot x(i) \qquad & \forall t \in \mathcal{T} \\
    y(t,j) \leq x(t) \qquad & \forall t \in \mathcal{T}, \forall j \in J \\
    x(t) \leq 1 \qquad & \forall t \in \mathcal{T}\\
    \sum_{t \in [t_1, t_2)} x(t) \ge \left\lceil \frac{\sum_{j \in J} q_j(I)}{g} \right\rceil  \qquad & I = [t_1, t_2), \forall t_1 \in \mathcal{T}, \forall t_2 \in \mathcal{T} 
    \end{aligned}
    }
    \end{equation*}
    \caption{C\u{a}linescu and Wang's linear program for active time scheduling~\cite{calinescu}}
    \label{figure:calinescu-lp}
\end{figure}


\begin{lemma} The linear program in Figure~\ref{figure:calinescu-lp} has an integrality gap of at least $3/2$ on nested instances.
\end{lemma}

\begin{proof}
The integrality gap instance consists of one long job $j_0$ with processing time $g$ and window $[0,2g)$, and $g$ groups of $g$ jobs. For $0\leq i < g$, the $i$-th group consists of $g$ jobs with unit processing time and window $[2i, 2i+2)$. 

Consider the following LP solution $(x,y)$: open each slot $t \in \mathcal{T} = [0, 2g)$ to an extent of $x(t) = (g+2)/2g$ for a total of $g+2$. For each $0 \leq i < g$, the LP schedules $1/2$ unit of $j_0$ and $1/2$ unit of each job $j$ in the $i$-th group in slots $2i$ and $2i+1$; that is, $y(2i, j_0) = y(2i+1, j_0) = 1/2$ and $y(2i, j) = y(2i+1, j) = 1/2$ for each job $j$ in group $i$. 

We now argue that $(x,y)$ satisfies the ceiling constraints of \cite{calinescu}'s LP; it is easy to check that the other constraints are satisfied. Consider an interval $I$. Since the long job $j_0$'s window is $[0,2g)$ and $j_0$ has length $g$, we have that $q_{j_0}(I) = 0$ if $|I| \leq g$ and $q_{j_0}(I) = |I| - g$ if $|I| > g$. This is because there are $2g - |I|$ slots outside of $I$. For a job $j$ in group $i$, since it has unit length, we have $q_j(I) = 1$ if $I$ contains its window $[2i, 2i+2)$ and $q_j(I) = 0$ otherwise. 

Combining the above, the LP constraint on interval $I$ is
\[\sum_{t \in I} x(t) \geq \left\lceil \frac{\max\{0, |I| - g\} + g|\{i \mid I \supseteq [2i, 2i+2)\}|}{g}\right\rceil.\]
The tightest constraints are when $I$ is the union of windows of consecutive groups. Thus, it suffices to argue that these constraints are satisfied. Suppose $I = [2i', 2(i'+k-1) + 2)$, i.e., it is the union of windows of $k$ consecutive groups. Note that $|I| = 2k$

If $k \leq g/2$, then the LP constraint on $I$ is $\sum_{t \in I} x(t) \geq k$. This is satisfied since $x_{2i} + x_{2i+1} = (g+2)/g$ for each group $i$. If $k > g/2$, then the LP constraint on $I$ is $\sum_{t \in I} x(t) \geq 1 + k$. This is also satisfied since $\sum_{t \in I} x(t) = k(g+2)/g = k + 2k/g > k + 1$. Thus, $(x,y)$ is a feasible solution to the LP that opens $g+2$ slots fractionally.

We claim that any integral solution $(x',y')$ needs to open at least $3g/2$ slots. Let $k$ be the number of groups $i$ such that $y'$ schedules at least one unit of the long job $j_0$ in the window $[2i, 2i+2)$. Consider the window $[2i,2i+2)$. Since there are $g$ unit jobs that need to be scheduled in $[2i,2i+2)$, $x'$ opens  two slots in the window if $y'$ schedules $j_0$ in the window and only one slot otherwise. Thus, $y'$ opens exactly $g + k$ slots. Since each window $[2i, 2i+2)$ has only two slots, and $j_0$ has length $g$, we have that $k \geq g/2$. Therefore, any integral solution needs to open at least $3g/2$ slots. Thus, the integrality gap of the LP is at least $\frac{3g}{2(g+2)}$ which converges to $3/2$ for large $g$.
\end{proof}

\section{NP-COMPLETENESS}
\label{sec:npc}
In this section, we show that the decision version of the nested active time problem is NP complete. Very recently, Sagnik~and~Manish~\cite{npcforgeneralcase} showed the general case is NP complete; unfortunately, their proof uses crossing intervals (i.e., intervals that overlap but neither is included in the other). Our proof reduces the nested active time problem to a new problem that we call \textbf{prefix sum cover}, which is related to the classic set cover problem.

\paragraph*{Prefix sum cover problem.}
For any pair of $d$-dimensional vectors $v = (v_1, v_2,..., v_d), w = (w_1, w_2, ..., w_d) \in R^d$, we say $v \prec w$ if and only if for all $j \in [1, d]$, $\sum_{i \leq j} v_i \geq \sum_{i \leq j} w_i$. In the prefix sum cover problem, we are given $n$ vectors $u_1, u_2, ...u_n \in N_{+}^d$, a target vector $v \in N^d$ and an integer number $k$, and we want to find $k$ vectors $u_{l_1}, u_{l_2}, ..., u_{l_k}$ such that $\sum_{i \leq k} u_{l_i} \prec v$.

Moreover, for the purposes of our reduction, we consider a restricted version of the problem. Let $W$ be the maximum scalar that appears in any of the vectors $u_1, ..., u_n$ and $v$. First, we require that both $d$ and $W$ be bounded by some polynomial of $n$. For a vector $w \in N^d$, let $[w]_j$ be its $j$-th dimension value.  Second, for each $i \in [1,n]$, we require that $[u_i]_1 \geq [u_i]_2 \geq ... \geq [u_i]_d$, and $[v]_1 \geq [v]_2 \geq ... \geq [v]_d$, i.e., all vectors are non-decreasing. Lastly, we require that all vectors are non-negative and integral.

\begin{proof} [NP Completeness of the prefix sum cover problem]
We will reduce set cover problem to the prefix sum cover problem. Recall $[v]_i$ is the $i$-th index value of vector $v$. Consider a set cover instance, $U$ is the universe containing $d$ elements, $S$ contains $n$ sets and $k$ is the target integer, the set cover problem is to find at most $k$ subsets from $S$ such that the union of those sets is the universe $U$. We could use a vector $u_i \in N^d$ to represent each set of $S$, for each index $j \in [1, d]$, if the set contains the $j$-th element, then we set the $j$-th value of $u_i$ to be 1. Otherwise, it will be 0. the target vector $v = \mathbf{1}^d$. Now the set cover problem is to find at most $k$ vectors $u_{l_1}, u_{l_2}, ..., u_{l_k}$ from $u_1, u_2, ..., u_n$ such that for each $j \in [1, d]$, $[\sum_{i \leq k} u_{l_i}]_j \geq [v]_j$. For technique problem, we will add $0$-th index to the vector and set it to be 0 for both $u$ and $v$. This won't affect our solution and this index is only helps for dealing with $1$-th index.

Next, we will transform all vectors $u_1, u_2,..., u_n$ to new vectors $u'_1, u'_2, ..., u'_n$. Specifically, for each vector $u_i = ([u_i]_1, [u_i]_2, ..., [u_i]_d)$, the new vector $u'_i = ([u'_i]_1, [u'_i]_2, ..., [u'_i]_d)$, where $[u'_i]_j = [u_i]_j - [u_i]_{j - 1} + 2 + d - j$ for all $j \in [1, d]$. Notice that $[u_i]_j$ is either 0 or 1, thus $[u'_i]_j \in [1, d + 2]$ . Now for the target vector $v$, we will set the new vector $v' = ([v']_1, [v']_2, ..., [v']_d)$ such that  $[v']_j = [v]_j - [v]_{j - 1} + 2k + k(d-j)$. Again, since $[v]_j, [v]_{j -1} \in [0, 1]$, we have $[v']_j \in [2k - 1, kd + k + 1]$. Thus, the maximum value in the new vectors is at most $kd + k + 1$, which is at most polynomial in $n$ and fits our requirement of prefix sum cover problem. Last, for the ordering requirement, we have $[u'_i]_j - [u'_i]_{j - 1} = [u_i]_{j - 2} - [u_i]_{j - 1} + 1 \geq 0$ and $[v']_j - [v']_{j-1} = [v]_{j-2} - [v]_{j-1} + k \geq 0$. Now we want to show the following if-and-only-if for the reduction.

\textbf{If part} If we have a solution $u_{l_1}, u_{l_2}, ..., u_{l_k}$ for the set cover problem. If the solution contains less than $k$ vectors, we could add some vectors to the solution until $k$ vectors, this doesn't change the solution, so we could assume we have $k$ vectors in the solution. Now, we want to show, the new vector $u'_{l_1}, u'_{l_2}, ..., u'_{l_k}$ is a solution for the partial sum problem. From set cover problem, we know $[\sum_{i \leq k} u_{l_i}]_j \geq [v]_j$, for $j \in [0, d]$. Our target is to show $\sum_{i' \leq j}\sum_{i \leq k}[{u'_{l_i}}]_{i'} \geq \sum_{i' \leq j} [v']_{i'}$, for $j \in [1, d]$.  Notice that 
\begin{equation*}
\begin{aligned}
    &\sum_{i' \leq j}\sum_{i \leq k}[{u'_{l_i}}]_{i'} - \sum_{i' \leq j} [v']_{i'} \\
    &= \sum_{i \leq k}\sum_{i' \leq j} ([u_{l_i}]_{i'} - [u_{l_i}]_{i' - 1} + 2 + d - i') - \\ &\sum_{i' \leq j} ([v]_{i'} - [v]_{i'-1} + 2k + k(d-j)) \\
    & = \sum_{i \leq k}([u_{l_i}]_j + 2j + \frac{(2d - 1 - j)j}{2}) -  ( [v]_{j} + 2jk + \frac{(2d - 1 - j)jk}{2}) \\
    & = \sum_{i \leq k} [u_{l_i}]_j - [v]_{j} = [\sum_{i \leq k} u_{l_i}]_j - [v]_{j} \geq 0
\end{aligned}
\end{equation*}
Thus, the new vectors are the solution for the partial problem.

\textbf{Only if} If we have a solution $u'_{l_1}, u'_{l_2}, ..., u'_{l_k}$ for the prefix sum cover problem. Again, if the solution contains less than $k$ vectors, we add some vectors to the solution. Notice that all number in the vector are non-negative, thus, the new solution is still feasible. Now, based on the above equation, the vector $u_{l_1}, u_{l_2}, ..., u_{l_k}$ is a solution for the set cover problem.

\end{proof}

\textbf{Remark:} Notice that the prefix sum cover problem is almost the same as set cover problem except for the "order" relation. We can think of the set cover problem as requiring that each dimension of the sum vector is greater than the target vector, while in the prefix sum cover problem, the requirement is "prefix sum".

\paragraph*{Reduction}
Now we will reduce the prefix sum cover problem to the active time problem. Let $(\{ u_1, u_2, \ldots, u_n\}, v, k)$ be the prefix sum cover instance. Our nested active time instance is defined by a set of jobs $J$ and it uses $p = dW$ machines. Our instance is made up of three kinds of jobs:
\begin{itemize}
    \item For each vector $u_i$, and each $w \in [2, W]$, we have $p - |\{j \in [1,d] \mid [u_i]_j \geq w\}|$ rigid unit length jobs, each with window consisting of a single slot $[(i-1)W+w - 1, (i-1)W + w]$.
     \item  For each vector $u_i$, we also have $\sum_{j \leq d} [u_i]_j - d$ flexible unit jobs with window $[(i-1)W, iW]$. 
     \item Finally, we have jobs that depend on the target vector. For each $j \in [1, d]$, we have a job with length $[v]_j$ and window $[0, nW]$.
\end{itemize}
We denote each of these sets of job with $S_1$ (rigid jobs), $S_2$ (flexible jobs associate with each $u_i$ vector), and $S_3$ (jobs associated with the target vector).

Let us try to schedule this instance, starting with $S_1$. Since the jobs in $S_1$ are rigid, we must open all slots in $[(i-1)W + 1, iW]$. Notice that each of these slots has at least $p - d$ jobs in $S_1$, so each of these time slots has at most $d$ unused machines after scheduling~$S_1$.

Next we will try to fit jobs from $S_2$ into $[(i-1)W, iW]$. Observe that the total capacity in the window $[(i-1)W+1, iW]$ is $p(W-1)$ and that the jobs from $S_1$ take up up $\sum_{w \in [2, W]} (p - |\{j \in [1,d] \mid [u_i]_j \geq w\}|)$ capacity. Further observe that \[\sum_{w \in [2, W]} (p - |\{j \in [1,d] \mid [u_i]_j \geq w\}|) + \sum_{j \leq d} [u_i]_j - d = p(W - 1).\] 
Therefore, if we do not open the slot $[(i-1)W, (i-1)W + 1]$, then the jobs from $S_1$ and $S_2$ will use up all of the available capacity in the time window $[(i-1)W, iW]$. This is important, as it means that we cannot schedule any job from $S_3$ in this window.


We say that the time slots $[(i-1)W, (i-1)W + 1]$ for $i \in [n]$ are \emph{special}. Since all non-special slots in $[0, nW]$ must be open, the problem boils down to opening as few special slots as possible to accommodate the jobs in $S_3$.

Suppose we open the special time slot $[(i-1)W, (i-1)W + 1]$.
We claim that all jobs in $S_2$ will be assigned to the special time slot. Indeed, even after all $S_2$ jobs are assigned to this slot, there are still $p - (\sum_{j \leq d} [u_i]_j - d) \geq d$ unused machines in it, while we can only have at most $d$ unused machines in each time slots in $[(i-1)W + 1, iW]$ after scheduling $S_1$.


Let \emph{configuration} be a sequence $(z_1, z_2, ..., z_M)$, where $z_i$ is the number of machines unused in time slot $[i - 1, i]$. Thus, once we have chosen which special slots to open, we get the configuration which tells us how many machines are left unused in each time slot. In the remainder of this section, we give an if-and-only-if condition on whether a configuration can fit all jobs from $S_3$.

Assume the machines are numbered from $1$ to $p$. For any given configuration, we can assume without loss of generality that if we have $z_t$ unused capacity at time slot $[t - 1, t]$ then machines $1$ through $z_t$ are unused; i.e, we always leaves smaller index machine unused. Let $e_j$ be the number of empty time slots at machine $j$. We give an if-and-only-if condition for the feasibility based on the $e_j$ values.

\begin{lemma}
\label{lem:reduction}
Given a configuration, let $e_j$ be the machine unused slot defined above and $J'$ be a set of $q \leq p$ jobs with no release time and due time constraint. Let $l_1 \geq l_2 \geq ...\geq l_q$ be the lengths of the jobs in $J$. The configuration can fit all jobs in $J'$ if and only if $\sum_{i \leq j} e_i \geq \sum_{i \leq j} l_i$ for all $j \in [1, q]$.
\end{lemma}
\begin{proof}
To identify each job, when we say the $i$-th job, we refer the job with length $l_i$.

\textbf{If part} Suppose  $\sum_{i \leq j} e_i \geq \sum_{i \leq j} l_i$ for all $j \in [1, q]$. Now, we prove the following statement by induction on $k$: if $\sum_{i \leq k} e_i \geq \sum_{i \leq k} l_i$ for any $k$, then we can fit jobs $l_1, l_2, ..., l_k$ into the first $k$ machines. The base case is $k = 1$: since $e_1 \geq l_1$, then we can fit $l_1$ into the first machine. Now we prove the inductive case. Suppose we can fit the first $k$ jobs into the first $k$ machines and we want to fit the first $k + 1$ jobs into the first $k + 1$ machines. Now, we first try to fit the first job to the first $k + 1$ machines and then use our induction. We fit the first job in following sequence: we first use machine $k + 1$ and if we cannot fit all of the first job in it, we use the machine $k$ and repeat this process, i,e, fit the first job by using machine with decreasing index. Notice that we use at most $e_1$ time slots to fit the first job, thus the above process will finally fit the first job inside but use some time slot of machines from $k + 1$ to $1$. Now, let $e'_1, e'_2, e'_3..., e'_{k + 1}$ be the machine unused time slot after we fit the first job inside. We know that $e'_{j} \geq e_{j + 1}$ for $j \in [1, k]$ since a job cannot use the same time slot twice. Another point is if $e'_{j} < e_j$, i.e, we use some time slot of machine $j$, then $\sum_{i > j}(e_i - e'_i) = e_{j+1}$. Now, notice that we have jobs $l_2, l_3,..., l_{k + 1}$, and the new time slot is $e'_1, e'_2, ... e'_k$, if we can show $\sum_{i \leq j} e'_i \geq \sum_{i \leq j} l_{i+1}$ for all $j \leq k$, then by induction, we can fit the second to the $(k+1)$-th jobs to the first $k$ machines. If $e'_j = e_j$, then we can say $e'_{j'} = e_{j'}$ for all $j' \leq j$, and thus $\sum_{i \leq j} e'_i = \sum_{i \leq j} e_i \geq \sum_{i \leq j}l_i \geq \sum_{i \leq j}l_{i+1}$. If $e'_j < e_j$, we have $l_1 = \sum_{i \leq k}(e_i - e'_i) = e_{j + 1} + \sum_{i \leq j}(e_i - e'_i)$, thus

\begin{equation*}
\begin{aligned}
    &\sum_{i\leq j+1}e_i \geq \sum_{i \leq j + 1}l_i = l_1 + \sum_{i \leq j}l_{i+1} = e_{j+1} + \sum_{i \leq j}(e_i - e'_i) + \sum_{i \leq j}l_{i+1}  \\
    &\Rightarrow \sum_{i\leq j}e_i - \sum_{i \leq j}(e_i - e'_i) \geq \sum_{i \leq j}l_{i+1} \Rightarrow \sum_{i \leq j} e'_i \geq \sum_{i \leq j} l_{i+1}
\end{aligned}
\end{equation*}

In either case, we know that the remaining jobs could be fitted into machines from 1 to $k$, thus we could fit all jobs if for all $j \in [1, p]$, $\sum_{i \leq j} e_j \geq \sum_{i \leq j} l_j$.

\textbf{Only if part} If for some $j \in [1, q]$, $\sum_{i \leq j} e_i < \sum_{i \leq j} l_i$, then the configuration is impossible to fit jobs $l_1, l_2, ..., l_j$. Consider a time slot $[t- 1, t]$, when we fit a job inside, we always use the machine with smallest index, if we could fit $l_1, l_2, ..., l_j$ into the time slot, then we will use machines with index at most $j$. If at any time slot, we use a machine with index greater than $j$, then we know we already use all machines from 1 to $j$, however, we have only $j$ jobs now. Thus, if we could fit the first $j$ jobs into the time slot, we can use the first $j$ machines to fit those jobs. However, for the first $j$ machines, the total capacity $\sum_{i \leq j} e_i$ is less than the length of all jobs, i.e, $\sum_{i \leq j} l_i$. Therefore, it is not possible to fit the first $j$ jobs into the configuration.
\end{proof}

Now we show how to apply it to the active time instance. We will set $J = S_3$ and $q = d$. For any interval $[(i-1)W, iW]$, let $e_{1, i}, e_{2, i}, ..., e_{d, i}$ be the unused time slot for machine $1$ to $d$ in this interval. If we close the special time slot $[(i-1)W, (i-1)W + 1]$, then there is no capacity left so $e_{1, i} = e_{2, i} = ... = e_{d, i} = 0$. If we open it, then $e_{j, i} = [u_i]_j$ , i.e.~the $j$-th machine will hold exactly $[u_i]_j$ unused time slots in the interval. Now the problem becomes we want to open $k$ special time slots such that the resulting configuration can fit all jobs from $S_3$. Lemma~\ref{lem:reduction} implies that it is equivalent to choosing $k$ vectors from $(e_{1, 1}, \ldots, e_{d, 1}) = u_1, ..., (e_{1, n}, \ldots, e_{d, n}) = u_n$ such that $\sum_{i\leq j} e_i \geq \sum_{i \leq j} [v]_i$ for every $j \in [1,d]$, which is exactly the definition of our prefix sum cover problem. Note that the ordering requirement comes from the fact we have ordering requirement in Lemma~\ref{lem:reduction} and the positiveness of $u$ comes from the fact that the machine from $1$ to $d$ has 1 unused space at time slot $[(i-1)W, (i-1)W + 1]$ if we open it. Since $d, W$ are all polynomial, the interval length and the machine number $p$ are also polynomial.

\bibliography{bib}


\end{document}